# Principles, Paradigms and the Future of UAV Drone Teams in Use for Engineering and Operation of Landscape-Scale Deployable Structures

John-Thones Amenyo

Department of Mathematics & Computer Science

York College, CUNY

jtamenyo@york.cuny.edu

**Abstract**

Models and Principles design principle models, techniques and tools

Drone fleets, with counts on the order of O(100) to O(1000), will play important and significant roles in the automation of the deployment, operation, maintenance and repair of ubiquitous and pervasive landscape scale elongated structures, with the longest linear spatial dimensions of O(1 mile) to O(10 mile). The organization of the drone team to support the task is considered as a digital platform, (specifically, a digital multi-sided platform). A computational thinking approach is used to engineer the architecture of the platform.

*Key words & phrases:* drone applications, landscape scale structures, design principles, computational thinking, modular engineering, robot-drone hybrids, tethered drones, teams of drones, fleets of drones.

**Introduction**

This paper addresses models and principles that can underlie the design and engineering of the technical architectures and infrastructures for using teams of drones to build and deploy, and ultimately maintain and repair *landscape scale structures*. A physical environmental structure is considered as being at landscape scale if at least one of its spatial dimensions, typically considered its length is of value O(1 mile) to O(10 mile). The deployable structure may be a vertical wall, a grounded linked chain or pipeline, or a suspended, aerial or airborne structure. The height dimension is on the order of O(100 ft) to O(1000 ft), and the width dimension is also on the order of O(300 ft) to O(1000 ft).

The landscape structure may in turn be just a section or segment of a much larger deployed structure, with length dimension reaching O(100 mile) to O(1000 mile).

*Problem Background*

Landscape scale structures are ubiquitous and pervasive. Several concrete examples include sections of utility networks for the transport, transmission, distribution and delivery of electric power, fuel, oil and gas, water, sewage, telephone and telecommunications, firefighting and fire control; as well as transport infrastructures: road systems, railways and canals.

Anticipating the near future, an interesting challenge is how teams of drones (flying robots), robots and robot-drone hybrids can be engineered, organized and deployed so as to completely automate the tasks of both (a) the construction and deployment and later (b) the operation, maintenance, repair, re-engineering and upgrade of these landscape structures. The problem of the cyber-technical architectures to support landscape scale use of fleets of drones have yet to be addressed systematically and in a thorough manner in the extant published literature.

The most serious challenge is the velocity or speed of rapid deployment work performance and achievement; that is, how quickly the deployment (or assembly, traversal, inspection, tour, sweep, scan) of the structure can be accomplished by the fleet of drones. For example, viable and effective deployments of landscape structures for fire-fighting and fire control and management on landscape scale have to be completed within time durations on the order of O(1 hour) to O(10 hour), in order to be considered useful and practical. A secondary but quite important cluster of challenges concern the exact and specific roles human workers will play in the engineering, installation, construction, (re-) configuration, deployment, operation and maintenance of the identified class of structures. Considerations of the Future of Work (FOW), (Balliester & ElSheikhi 2018), suggest that the task niche can be and will be completely automated, at all layers and scales, both for operational and supervisory or executive tasks, with humans playing very limited and minimal roles. This will be particularly relevant for truly "dull, dirty and dangerous" tasks in the niche domain.

The approach uses an integrated set of computational metaphors to design and engineer the needed architecture.

*Related Work*

To the best of the author's knowledge, this is one of the first studies that specifically and systematically addresses the organization and engineering of the viable and effective architectures and infrastructures in the application area of landscape scale structures and networks. Currently, the main uses of drones in real-estate, utility networks, transport, fire-fighting, are to take photos, pictures, videos for asset or structure evaluation, assessment, inspection, surveillance, monitoring and tracking, data gathering and acquisition, explorations and surveys, (Ascend Technology 2018, Shakhatreh 2018, USAID 2017, Beloev 2016). Drones have also been used for search-and-rescue missions, following natural disasters. Drones have also been used to drop fire retardants, although not on a large scale.

**Materials and Methods**

The team of drones for use in the application domain, is considered as an organized society of functionally specialized entities. Computationally, it is regarded as a coordinated Macro-Project Platform. Thus, the task architecture problem is transformed into the engineering of the architecture of a coordinated platform. The landscape scale structure deployment task is taken as requiring O(10) to O(100) team of drone machines. A drone can be embodied as a multi-copter or UAV, although several additional embodiments are also taken into account, including aerostat, dirigibles, balloons, ground robots, ocean surface robots, underwater robots, robot-drone hybrids and chimeras. Of necessity, the drones need to be autonomous, and driven, controlled, coordinated and managed by stored-program algorithms; in the advanced forms by AI and machine intelligence based algorithms.

The collective of drones is referred to as a team (or fleet, group, gang, troop, project, platoon, legion, ensemble), because the terms such as swarm or ensemble would seem to indicate collection counts on the order of O(1000) to O($10^6$) individual drones.

The most serious practical limitation to field application of current (commercial and research) drones is availability of onboard power. ICE (internal combustion engine) and EV-ICE hybrid drones have been researched and demo'ed, but are not currently commercialized. Therefore, the platform architecture design and engineering model assumes that any drone in the team that requires power operational duration of more than O(1 hour), will be deployed as a tethered drone. The tether cables to such a drone will carry electric power, and may carry communications and control signals, sensor and IOT data, as well as telemetry and telepresence data. Thus, aside from the issues of managing coordinating individual and sub-groups of tethered drones, because of possible entanglement of tethering, the issue of power supply and availability to long duration drones is obviated in the chosen application domain.

The primary core requirements (system qualities or *iLities*) for the targeted application domain are:

- Functionality
- Productivity (machine or drone)
- Performance
- Cost

The landscape scale structures are deployed in the geo-spatial context of areas, zones or sites of:

- Field Operations: where the deployable structures are actually co-located
- Field Staging Areas: forward or front-end command stations
- Central Command Centers: Back-end Situation HQ
- Off-site Supply Stations

This geographical division gives rise to a lineup of drones playing specialized roles:

- Supply Station drones
- Central Command Center drones
- Field Staging Area drones
- Field Operation drones
- Field-to-Offsite Ferry and Taxi drones
- Telepresence drones, Tele-expert drones
- Other miscellaneous drones

All the drones to be deployed are expected to be provisioned to have the following technical characteristics, affordances and subsystems, (Hristov, Zahariev & Beloev 2016):

- Modular and reconfigurable in construction: configured by combining several standard building blocks – to facilitate reusability, re-configurability, flexibility, (field) re-programmability and adaptability.
- Ducted or being spherical, ball, encapsulated, enveloped – for safety of human and other machine co-workers.
- Autonomous and (human) driverless.
- Speakers & Microphones – for interactions with human co-workers
- Sensors – sense ambient environment, surroundings, external context; hence, incorporate IOT, dense sensor network; smart devices, intelligent components, that can provide self-awareness, other-awareness, environmental-awareness, ambient-awareness, sensitivity and responsiveness.
- Attachable and detachable augmentations and auxiliaries:

a. Robotic arms, cables, rope wires, tethers for payload handling
   b. Cargo and payload transport and delivery frameworks, containers, sleeves, cages, mesh nets
   c. Lights, for night time operation
   d. Devices for self-security, self-protection
   e. Devices for fault-tolerance of emergencies, disasters and crises (EDC/FT)

Individual UAV drones are assumed to have functionally specialized roles, by adopting and incorporating subsets of the following basic task affordances and capabilities:

> Photography; Imaging; Telemetry; Close-up inspection; Inspection reporting; Survey; Close-up Search via Sweeps and Scans; payload-cargo-container handling; Tools, instruments, equipment Handling; (sub-) Structure Assembly; Vertical or elevation (up and down) movement of payload objects, articles, tools, equipment and structure elements; Lateral or horizontal movements of payload objects; and sweep-scanning traversal of deployed structures; Payload-Package Delivery.

The main Operational Workflow chain:

Central HQ -> Supply Station -> Field Staging Area -> Field Operations -> Field Staging Area -> Supply Station -> Central HQ.

The complete Operational Workflow schema is lattice structured directed graph.

**Results and Discussion**

*Results*

Several key design and engineering principles are used to formulate the desired architecture and infrastructure. The topmost principle is to use the metaphor of Colony Organism, (for example, a bureaucratic organization, navy or armada), to model and represent the overall architecture. This means the overall deployed system is regarded as embodiments, mechanizations and digitization of (multiple) co-existing and integrated separations of concerns, aspects and roles (SOCAR), a platform of an integrated collection or society of agencies (Minsky).

The second most important principle is the use of Computational Thinking embodied Modular Programming, (much like visual programming via building blocks of the Scratch language), in all aspects of the architectural specification and design.

The needed Landscape Scale Work Architecture (Work Architecture) is designed using the following algorithmic steps:

(1) The (field deployable, landscape scale) spatial structure is Compartmentalized into domains (or compartments, zones, sections, areas, segments). The tasks required to support configuration and deployment, operational management, maintenance and repair, upgrades and re-engineering of the field structure are allocated to spatial domains, to form task compartments.

(2) Each task compartment is treated as an independent, isolated, segregated, autonomous system, in its own right, whose internal operations are not directly affected by the internal operations of other task compartments. The only indirect inter-domain inter-dependencies, linkage and coupling occur at the interfaces and connections boundaries of the task compartments. The above engineering design is equivalent to the Embarrassingly Parallel computation pattern, of data parallelism and result parallelism (Gelernter, Carriero), function style programming (Iverson, Backus, More) and algorithmic skeleton patterns (Cole, Flynn). Such patterns can be termed MIMD (MPMD) and SIMD (SPMD), depending on whether the same or different task instructions are to be executed, at each task micro-step, stage or phase.

(3) There is a need for coordination, communication, control, cybernetics choreography (C*) to address concerns and issues of co-dependence, concurrency, shared resource sharing contention and conflicts, (localized bulk) synchronization (Valiant), due to common, shared areas of intersections, interfaces, inter-connections of the task compartments. Two architectural options are available to handle the C* issues of common intersections areas of task compartments, (a) use of a third-party, broker, middleman or (nano- or micro-) multi-sided platform, which is a multi-agent team of dedicated and specialized intelligent cognitive agencies (ICA), or (b) the dynamic election of one compartment in an intersection relationship, as the Leader to handle the C* issues, while the rest of the compartments in the relationship are Followers, dictated to and modulated by the leader. For example, a C* handler or Leader can issue work orders, commands and instructions to the multiple Followers, in an intersection relationship. Inter-communications follow the processes of Stigmergy and the MapReduce computational patterns.

(4) In each task compartment, there are Technology Stacks to support embodiments of both conventional engineering attributes or qualities, the so-called *iLities*, such as functionality, reliability, survivability, security, efficiency, maintainability, adaptability, etc., as well as self-* and autonomics qualities and properties, self-repair, self-awareness, etc.

(5) Each of the multiple engineering attributes gives rise a multi-scale goal-agenda-task-behavior-operation-action tree or hierarchy.

(6) The Work Architecture assigns a drone subgroup to each (sub-) goal in the multiple goal trees. The drone workgroup serves as the society of agencies that can be used to realize and achieve the associated goal.

(7) In most practical deployments, there will not be enough concrete, physical drones available that can be allocated to each (sub-) goal. Hence, the Work Architecture treats each goal drone subgroup as a society or collective of virtual drones, (for example, each virtual drone is a digital twin of a physical drone).

(8) The Work Architecture includes a Work Operating System that coordinates how the virtual drones can use various multiplexing, multi-access schemes to use the physical drones, to accomplish their assigned goals-tasks-behaviors-actions.

(9) At the lowest level or finest scale, the Work Architecture is designed to operate dynamically as an alternating bivariate process. Time periods (E*-phases) when task compartments operate as embarrassingly parallel sub-systems, are alternate with C*-phases, when there is coordination, synchronization of co-dependencies, and resolution of (physical shared resource) concurrency related multi-access conflicts and contentions, (Fig. 1).

ig

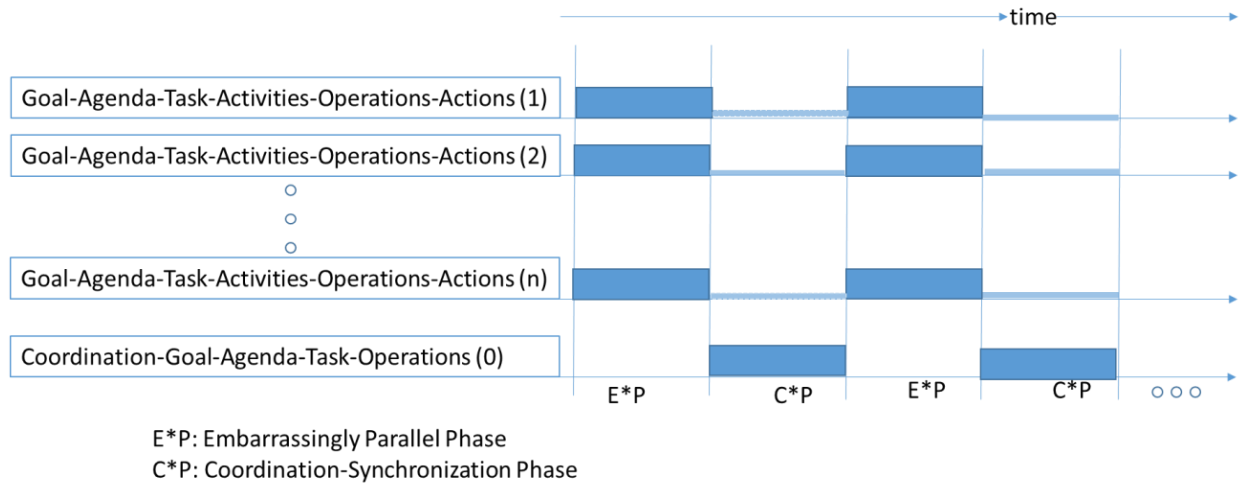

Fig. (1): Work Architecture Task Alternating Process

**Summary and Conclusions**

Life cycle development, operation and support of landscape structures is one niche of drone applications that is pregnant with lots of opportunities. A serious challenge is the design and engineering of the architecture and infrastructure of the platforms that can be used to support such applications. A computational thinking approach that emphasizes modular engineering using architectural building blocks can be used to specify and construct such landscape scale automation, control, coordination and cybernetic architectures.

One immediate direction of future research is to build prototypes of the architecture design described in the paper, for field deployment and testing. A sub-area of particular interest is the engineering of digital twins of operational systems and platforms to support the deployment, operation and management of landscape-scale structures.

Another intriguing possibility is to combine and integrate multiple landscape scale designs to create architectures and infrastructures of systems that can be used to control of tropical storms, such as hurricanes and typhoons.